\documentclass{ws-ijbc}
\usepackage{ws-rotating}     
\usepackage{graphicx}
\usepackage{epstopdf}
\begin{document}
\catchline{}{}{}{}{} 

\markboth{Katsanikas et al.}{Detection of Dynamical Matching in a Caldera Hamiltonian System using Lagrangian Descriptors}

\title{Detection of Dynamical Matching in a Caldera \\ Hamiltonian System using Lagrangian Descriptors}

\author{M. Katsanikas} 
\address{School of Mathematics, University of Bristol, \\ Fry Building, Woodland Road, Bristol, BS8 1UG, UK \\ matthaios.katsanikas@bristol.ac.uk}

\author{V\'ictor J. Garc\'ia-Garrido}
\address{Departamento de F\'isica y Matem\'aticas, Universidad de Alcal\'a, \\ Alcal\'a de Henares, 28871, Spain. \\ vjose.garcia@uah.es}

\author{S. Wiggins}
\address{School of Mathematics, University of Bristol, \\ Fry Building, Woodland Road, Bristol, BS8 1UG, UK \\  s.wiggins@bristol.ac.uk}

\maketitle

\begin{history}
\received{(to be inserted by publisher)}
\end{history}

\maketitle

\begin{abstract}
	
The goal of this paper is to apply the method of Lagrangian descriptors to reveal the phase space mechanism by which a Caldera-type potential energy surface (PES) exhibits the dynamical matching phenomenon. Using this technique, we can easily establish that the non-existence of dynamical matching is a consequence of  heteroclinic connections  between the unstable manifolds of the unstable periodic orbits (UPOs) of the upper index-1 saddles (entrance channels to the Caldera) and the stable manifolds of the family of UPOs of the central minimum of the Caldera, resulting in the temporary trapping of trajectories. Moreover, dynamical matching will occur when there is no heteroclinic connection, which allows trajectories to enter and exit the Caldera without interacting with the shallow region of the central minimum. Knowledge of this phase space mechanism is relevant because it allows us to effectively predict the existence, and non-existence, of dynamical matching. In this work we explore a stretched Caldera potential by means of Lagrangian descriptors, allowing us to accurately compute the critical value for the stretching parameter for which dynamical matching behavior occurs in the system. This approach is shown to provide a tremendous advantage for exploring this mechanism in comparison to other methods from nonlinear dynamics that use phase space dividing surfaces. 

\end{abstract}

\keywords{Chemical reaction dynamics; Phase space transport; Hamiltonian systems; Lagrangian descriptors; Periodic orbits; Invariant manifolds; Symmetry; Caldera potential; Poincar{\'e} sections}

\section{Introduction}
\label{intro}
Dynamical matching is an interesting mechanism originally proposed in \cite{carpenter1985,carpenter1995} that arises in Caldera-type potential energy surfaces (PES). These potentials are relevant in chemistry since they provide good approximations for the description of many organic chemical reactions, such as those that occur in the vinylcyclopropane-cyclopentene rearrangement \cite{baldwin2003,gold1988}, the stereomutation of cyclopropane \cite{doubleday1997}, the degenerate rearrangement of bicyclo[3.1.0]hex-2-ene \cite{doubleday1999,doubleday2006} or that of 5-methylenebicyclo[2.1.0]pentane \cite{reyes2002}. The potential energy surface of a Caldera is similar to that of a collapsed region of an erupted volcano. It is characterized by a shallow potential well region (a central minimum) surrounded by four entrance/exit channels mediated by index-1 saddles. Two of these saddles have low energy values and correspond to the formation of chemical products, while the other two are higher in energy and represent reactants. 

Broadly speaking, trajectories in Caldera type PES exhibit two distinct types of dynamical behavior. The first kind is the trapping of trajectories in the central minimum area of the Caldera, and the other type is dynamical matching. Examples of the behavior of these types of trajectories for the type of Caldera PES studied in this paper were described in \cite{collins2014}. In the first case, trajectories that have initial conditions on the dividing surfaces of the unstable periodic orbits (UPOs) of the upper index-1 saddles enter the central area of the Caldera and become temporarily trapped as a result of the interaction between the invariant manifolds of the UPOs that exist in the central area of the Caldera with those of the unstable periodic orbits of the index-1 saddles. This is studied in  \cite{katsanikas2018}. Eventually, these trajectories will exit the Caldera through any channel corresponding to the four index-1 saddles surrounding the central area. As we will show in this work, trapping of trajectories, i.e. non-existence of dynamical matching, is a consequence of  heteroclinic connections between the stable manifolds of the family of UPOs in the central minimum of the Caldera and the unstable manifolds of the UPO of the upper index-1 saddles. 

The second type of trajectory behavior is dynamical matching, for which trajectories with initial conditions on the dividing surfaces of the UPOs of the upper index-1 saddles go straight across the Caldera and exit via the opposite lower index-1 saddles. This was considered in \cite{katsanikas2018}. The understanding of this mechanism is very important  for Caldera PESs with reflectional symmetry about the $y$-axis (which is what we consider in this paper)  since for such PESs statistical theories would predict that reactive trajectories exit with equal probability through the two channels of the lower index-1 saddles. However, chemical systems whose energy landscape possesses caldera intermediate regions on their PES almost never exhibit the expected symmetry in the product formation ratio. For this reason this mechanism must be understood from a phase space perspective. 

Dynamical matching  can be viewed as an expression of  momentum conservation and Newton’s first law of motion. It is manifested by a trajectory entering the Caldera from a channel corresponding to a high energy index-1 saddle (reactant). In the relatively flat region of the caldera it experiences little force, and it exits through the diametrically opposing low energy index-1 saddle (product). As  a result, this mechanism plays an important role in determining  the outcome of the chemical reaction. However, not all trajectories entering the caldera behave in this fashion. Some trajectories may interact with the shallow potential well region and become temporarily trapped. This can play a significant role in how they exit from the well.

In our previous study of dynamical matching for Caldera PES described in  \cite{katsanikas2018} we used the method of Poincar{\'e} sections to understand that dynamical matching is a consequence of the 
non-existence of interaction between the unstable invariant manifolds of the UPOs associated with the upper index-1 saddles and the manifolds from the central minimum of the Caldera. We also investigated in \cite{katsanikas2019} the conditions for the non-existence of dynamical matching in cases where we stretched the PES in the $x$-direction. In this case, the distance in the $x$-direction between the saddles and the central minimum increases as we decrease the stretching parameter. We found that there existed a critical value of the stretching parameter for which the system does not exhibit dynamical matching. At this critical value, the invariant manifolds of the UPOs associated with the upper index-1 saddles begin to interact with the central area of the Caldera, and trajectories become temporally trapped. We showed that this results from the decrease of the H{\'e}non stability parameter of the UPOs of the upper index-1 saddles that is responsible for the focusing of the unstable manifolds of the UPOs towards the central area of the Caldera \cite{katsanikas2019}. 

\cite{katsanikas2018,katsanikas2019} used the following methods to reveal and analyze the phase space structure: 

\begin{enumerate} 

\item Computation of periodic orbits using classical methods. In particular, it was noted that in Caldera-type Hamiltonian systems it is difficult to compute the Lyapunov families of UPOs of the index-1 upper saddles, since the system has distinct escape routes leading to  non-convergence of the methods in a reasonable computational time. 

\item Computation of periodic orbit dividing surfaces associated with relevant UPOs.

\item Computation of selected Poincar{\'e} sections.

\item Computation of the invariant manifolds of the UPOs on Poincar{\'e} sections.

\end{enumerate}

\noindent
In this paper we show how the method of Lagrangian descriptors can be used to achieve each of these steps with significant computational efficiency, both in implementation and time.

The outline of this paper is as follows. In section \ref{sec.1} we briefly describe the Caldera Hamiltonian system for which we analyze the dynamical matching mechanism. Section \ref{sec.1a} is devoted to introducing the method of Lagrangian descriptors and how it can be applied to reveal the geometrical template of invariant manifolds in the high-dimensional phase space of Hamiltonian systems. In section \ref{sec.2} we present the results of this work on how to detect the dynamical matching phenomenon using Lagrangian descriptors. Finally, in the last section we discuss the conclusions.   

\section{The Hamiltonian Model}
\label{sec.1}

\begin{figure}[htbp]
	\begin{center}
		\includegraphics[scale=0.26]{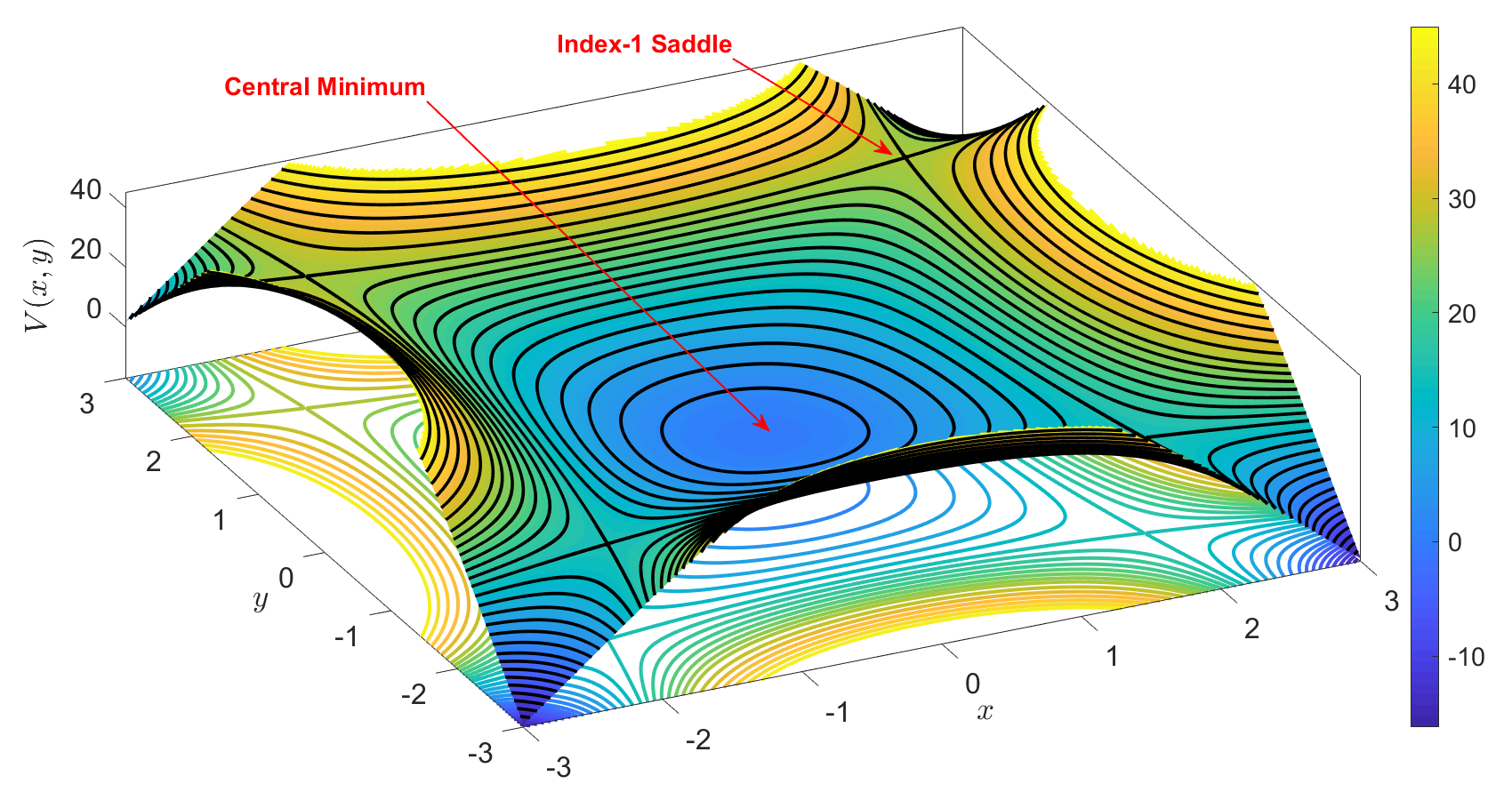}
	\end{center}
	\caption{Caldera potential energy surface given in Eq. (\ref{eq1}) for the model parameters $c_1 = 5$, $c_2 = 3$, $c_3 = -3/10$ and $\lambda = 1$.}
	\label{caldera_pes}
\end{figure}

\begin{figure}[htbp]
	\begin{center}
		A)\includegraphics[scale=0.48]{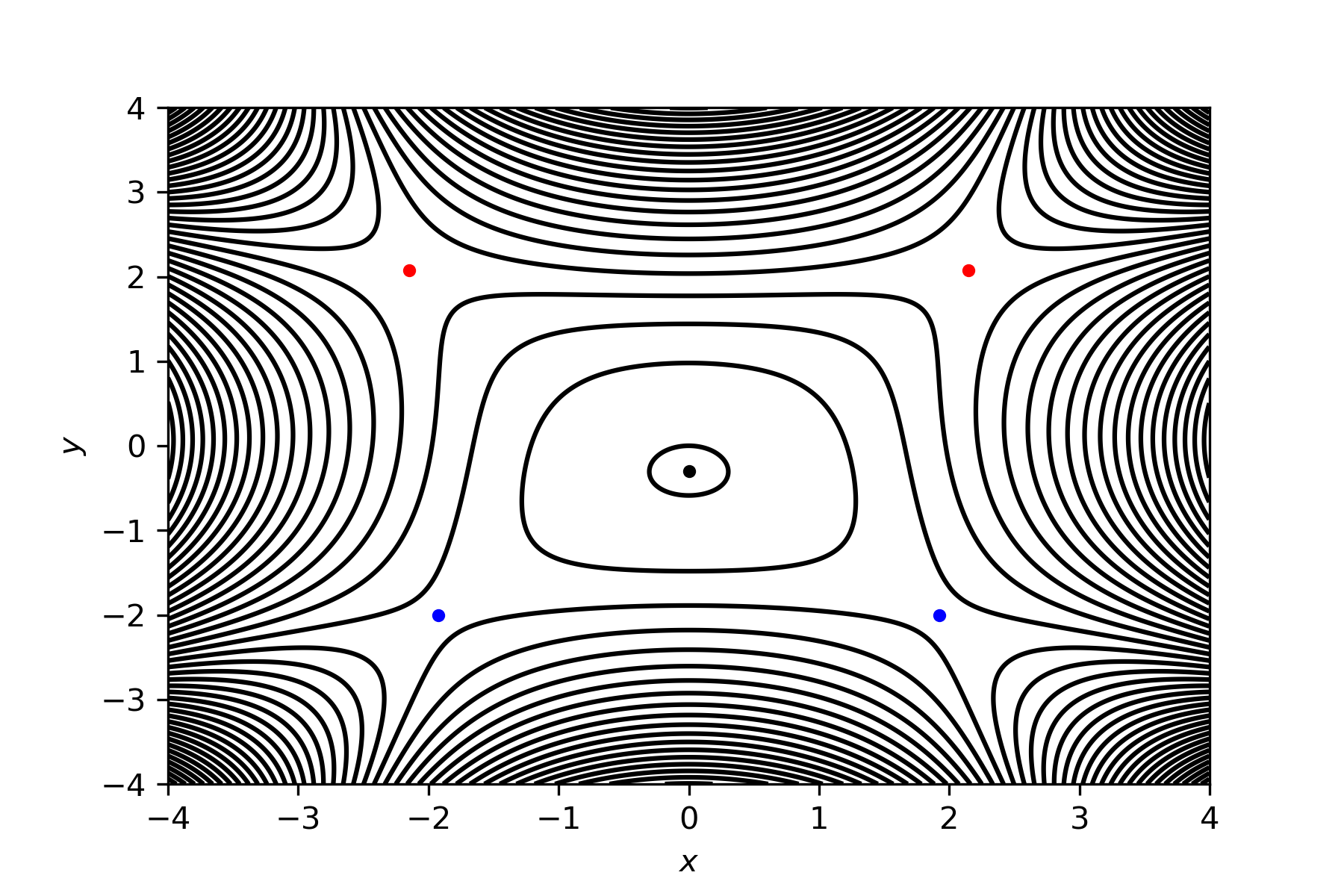}
		B)\includegraphics[scale=0.48]{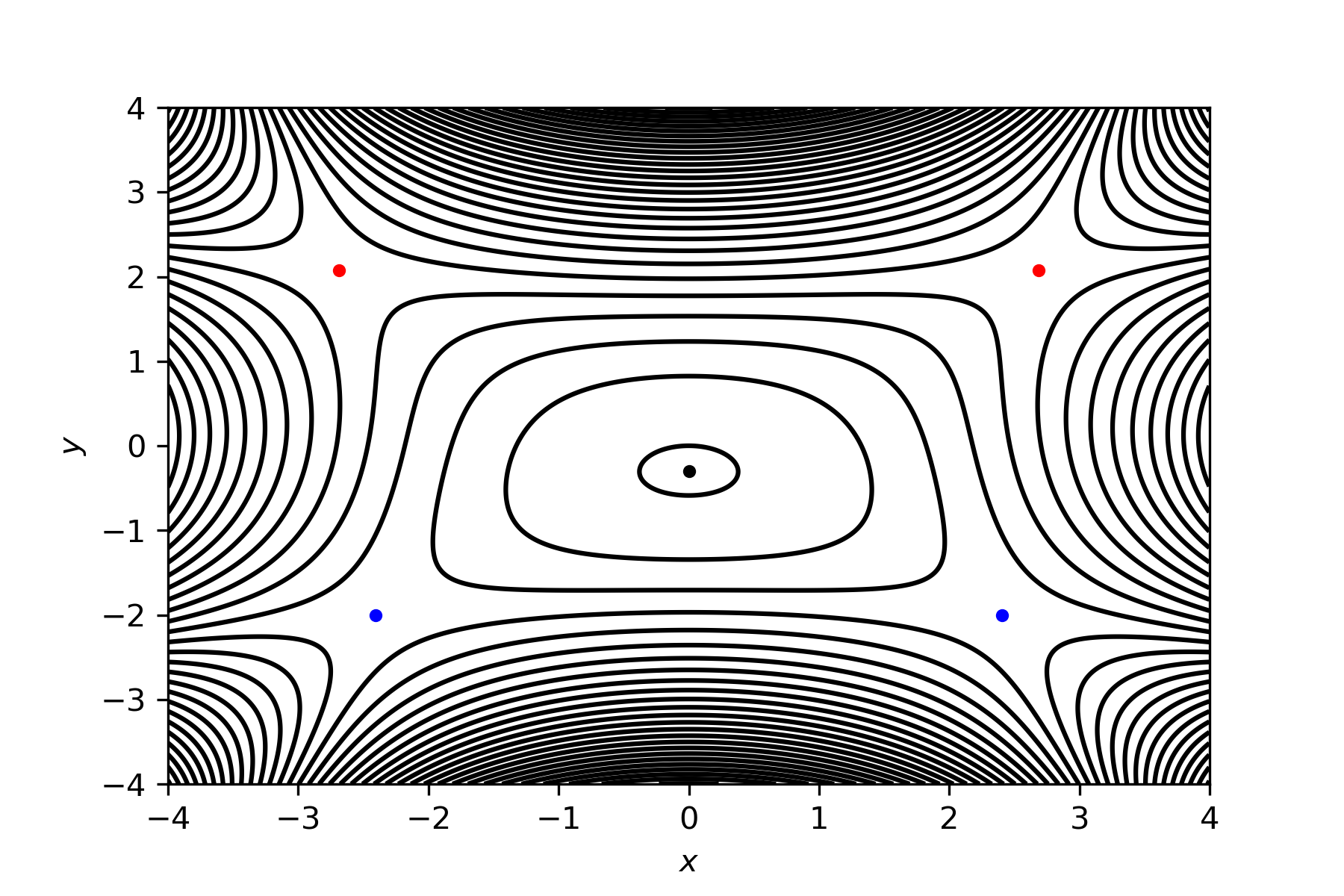}
		C)\includegraphics[scale=0.48]{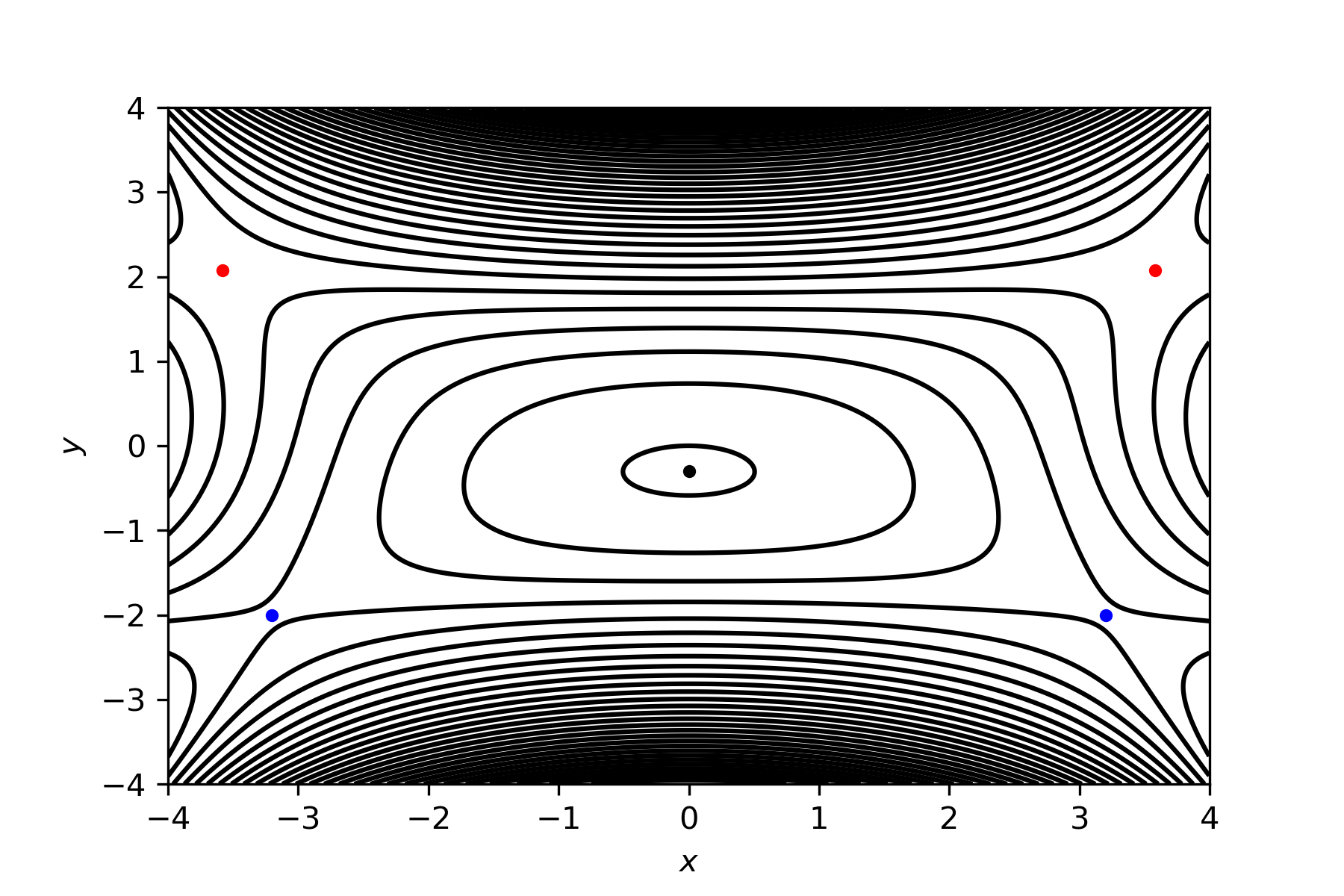}
		D)\includegraphics[scale=0.48]{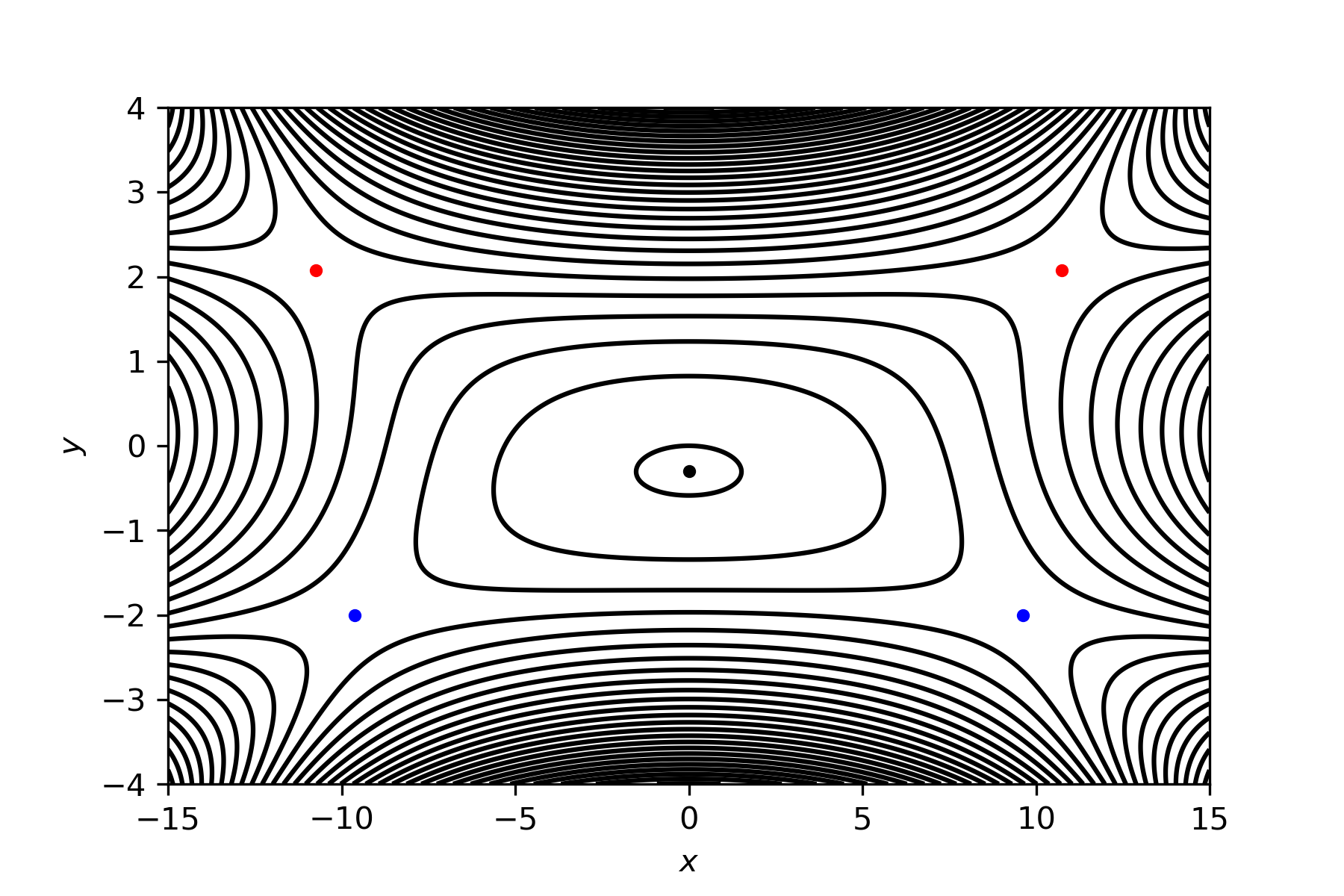}
	\end{center}
	\caption{The stable stationary point in the center area (depicted by a black point), the upper saddles (depicted by red points), the lower saddles (depicted by blue points) and the equipotential contours for the stretching parameter: A) $\lambda = 1$; B) $\lambda = 0.8$; C) $\lambda = 0.6$ and D) $\lambda = 0.2$.}
	\label{equi}
\end{figure}

In this section we present the Caldera PES that we have used in order to analyze the phase space structures responsible for the dynamical matching mechanism. The model PES that we consider, which has been addressed in previous works, see e.g. \cite{collins2014,katsanikas2018,katsanikas2019}, has a central minimum and four index-1 saddles around it. Two of these saddles have high energy values and the other two are lower in energy. Therefore, the regions about the index-1 saddles allow entrance and exit to and from the central area of the Caldera. In particular, we study a stretched version of the Caldera potential in the $x$ degree of freedom, in the form:
\begin{equation}
V(x,y) = c_1 \left(y^2 + (\lambda x)^2\right) + c_2 \, y - c_3 \left((\lambda x)^4 + y^4 - 6 \, (\lambda x)^2 y^2\right)
\label{eq1}
\end{equation}

\noindent where the model parameters used in this work are $c_1 = 5$, $c_2 = 3$, $c_3 = -3/10$ and $0 < \lambda \leq 1$ (the stretching parameter). The classical symmetric caldera PES \cite{collins2014,katsanikas2018} corresponds to $\lambda = 1$ and is shown in Fig. \ref{caldera_pes}. We depict in Fig. \ref{equi} the contours and the equilibrium points of the potential for different values of $\lambda$, for example $\lambda=1$, $\lambda=0.8$, $\lambda=0.6$ and $\lambda=0.2$. We also compile in Table \ref{tab:ta08} the positions and energies of the upper index-1 saddles for different values of  $\lambda$. We observe that the positions of the index-1 saddles move away from the center of the Caldera as we decrease the parameter $\lambda$.  The position of the central minimum is $(x,y) = (0,-0.297)$ with energy $E = -0.448$ for all values of the stretching parameter $\lambda$.

The Hamiltonian with two degrees of freedom is defined as the sum of kinetic plus potential energy:
\begin{equation}
H(x,y,p_x,p_y) = \frac{p_x^2}{2m_x} + \frac{p_y^2}{2m_y} + V(x,y)
\label{eq2}
\end{equation}
where $V(x,y)$ is the Caldera PES in Eq. \eqref{eq1}, and $m_x$, $m_y$ are the masses of the $x$ and $y$ DoF respectively. We denote the numerical value of the Hamiltonian as energy $E$. In this work we take $m_x = m_y =1$, and Hamilton's equations of motion are given by:
\begin{equation}
\begin{cases}
\dot x = \dfrac{\partial H} {\partial p_x} = \dfrac{p_x}{m_x} \\[.4cm]
\dot y = \dfrac{\partial H} {\partial p_y} = \dfrac{p_y}{m_y} \\[.4cm]
\dot p_x = -\dfrac{\partial H} {\partial x} = 2 \lambda \, (\lambda x) \left[2c_3 \left((\lambda x)^2 - 3 y^2 \right) - c_1 \right] \\[.4cm]
\dot p_y = -\dfrac {\partial H} {\partial y} = 2 y \left[ 2 c_3 \left(y^2 - 3 (\lambda x)^2\right) - c_1 \right] - c_2
\end{cases}
\label{eq3}
\end{equation}

\begin{table}[htbp]	
	\tbl{The upper index-1 saddles of the potential given in Eq. \eqref{eq1} ("RH" and "LH" are the abbreviations for right hand and left hand respectively) for different values of $\lambda$. The energy for all the cases is $E = 27.0123$.} {
		\begin{tabular}{| l c c c |}
			\hline
			Critical point & x & y & $\lambda$ \\
			\hline
			Upper LH index-1 saddle  &-2.149   & 2.0778 & 1 \\
			Upper RH index-1 saddle  &2.149    &  2.0778 & 1\\
			Upper LH index-1 saddle  &-2.6862  & 2.0778 & 0.8 \\
			Upper RH index-1 saddle  &2.6862  &  2.0778 & 0.8\\
			Upper LH index-1 saddle  &-3.5815  & 2.0778 & 0.6 \\
			Upper RH index-1 saddle  &3.5815  &  2.0778 & 0.6 \\
			Upper LH index-1 saddle  &-10.7446 & 2.0778 & 0.2 \\
			Upper RH index-1 saddle  &10.7446 &  2.0778 & 0.2 \\
			\hline
 \end{tabular} \label{tab:ta08} } 
\end{table}


\section{Lagrangian Descriptors}
\label{sec.1a}

The method of Lagrangian descriptors (LDs) is a trajectory-based scalar diagnostic tool that has been developed in the nonlinear dynamics literature to explore the geometrical template of phase space structures that characterizes qualitatively distinct dynamical behavior. This technique was originally introduced a decade ago in \cite{madrid2009} for the location of \textit{Distinguished Hyperbolic Trajectories}, and was defined by means of computing the arclength of particle trajectories as they evolve forward and backward in time \cite{mancho2013lagrangian}. The method was originally applied to study transport and mixing mechanisms in geophysical flows \cite{mendoza2010}. Recently, the technique has received recognition in the field of Chemistry, in particular in the area of Transition State Theory (see e.g. \cite{craven2015lagrangian,craven2016deconstructing,craven2017lagrangian}), where the computation of chemical reaction rates relies on the knowledge of the phase space structures that separate reactants from products. Therefore, the use of mathematical techniques that have the capability of detecting  high-dimensional phase space structures that occur in Hamiltonian systems, such as normally hyperbolic invariant manifolds (NHIMs) and their stable and unstable manifolds, is of great interest and utility. One of the biggest challenges when exploring the high-dimensional phase space of a dynamical system is to interpret the dynamical behavior of ensembles of initial conditions, and to  recover from the evolution of their trajectories the underlying geometrical phase space structures that govern the  dynamics. The problem that arises is that classical techniques rely on following the location of the trajectories of initial conditions that start nearby, and in a high-dimensional phase space, trajectories might get ``lost'' with respect to each other very quickly. The method of Lagrangian descriptors provides a radically different approach that resolves this issue, as it focuses on integrating a positive scalar function along the trajectory of any initial condition of the system instead of tracking their phase space location. This is probably one of the key ideas behind the success of this technique, as the phase space geometry is concealed in the initial conditions themselves.

In the framework of Hamiltonian systems it has been mathematically proven that LDs detect the geometrical phase space structures responsible for transition dynamics through index-1 saddles \cite{naik2019a}, and numerical studies have been carried out to analyze escaping dynamics on open PESs \cite{demian2017,naik2019b,GG2019}. The methodology offered by LDs has been shown to have many advantages with respect to other nonlinear dynamics tools. For instance, it is straightforward to implement and computationally inexpensive when applied to systems with two or three DoF. But probably the most important feature of this tool is that it allows to produce a complete and detailed geometrical \textit{phase space tomography} in high dimensions by means of using low-dimensional phase space probes to extract the intersections of the phase space invariant manifolds with these slices \cite{demian2017,naik2019a,naik2019b,GG2019}. 

Consider a dynamical system with general time-dependence in the form:
\begin{equation}
\dfrac{d\mathbf{x}}{dt} = \mathbf{v}(\mathbf{x},t) \;,\quad \mathbf{x} \in \mathbb{R}^{n} \;,\; t \in \mathbb{R} \;,
\label{gtp_dynSys}
\end{equation}

\noindent
where the vector field $\mathbf{v}(\mathbf{x},t) \in C^{r}  (r \geq 1)$ in $\mathbf{x}$ and continuous in time. In this work, this system is given by Hamilton's equations for the Caldera PES, see Eq. \eqref{eq3}. In order to explore the phase space structures of this dynamical system we have used a modified version of the $p$-norm definition of Lagrangian descriptors that relies on variable time integration. The reason for doing so is that, since the Caldera PES is an open potential, trajectories can escape to infinity at an increasing rate, and this issue may cause problems when computing LDs. Take an initial condition $\mathbf{x}_0 = \mathbf{x}(t_0)$ and a fixed integration time $\tau > 0$, the $p$-norm LD introduced in \cite{lopesino2017} is defined as follows:
\begin{equation}
M_p(\mathbf{x}_{0},t_0,\tau) = \int^{t_0+\tau}_{t_0-\tau} \, \sum_{i=1}^{n} |v_{i}(\mathbf{x}(t;\mathbf{x}_0),t)|^p \; dt = M_p^{(b)}(\mathbf{x}_{0},t_0,\tau) + M_p^{(f)}(\mathbf{x}_{0},t_0,\tau) \;,\quad p \in (0,1] \; .
 \label{Mp_function}
\end{equation}

\noindent
where $M_p^{(b)}$ and $M_p^{(f)}$ represent, respectively, backward and forward integration of initial conditions starting at time $t_0$, that is:
\begin{equation}
M_p^{(b)}(\mathbf{x}_{0},t_0,\tau) = \int^{t_0}_{t_0-\tau} \sum_{i=1}^{n} |v_{i}(\mathbf{x}(t;\mathbf{x}_0),t)|^p \; dt \quad,\quad M_p^{(f)}(\mathbf{x}_{0},t_0,\tau) = \int^{t_0+\tau}_{t_0} \sum_{i=1}^{n} |v_{i}(\mathbf{x}(t;\mathbf{x}_0),t)|^p \; dt
\end{equation}

\noindent
In particular, we have chosen for this work $p = 1/2$. At this point, it is important to highlight that with this definition of LDs one can mathematically prove that NHIMs and their stable and unstable manifolds are detected as singularities of the $M_p$ scalar field, that is, points at which the function is non-differentiable and thus its gradient takes very large values \cite{lopesino2017,demian2017,naik2019a}. Moreover, it has been shown that,
\begin{equation}
\mathcal{W}^u(\mathbf{x}_{0},t_0) = \textrm{argmin } M_p^{(b)}(\mathbf{x}_{0},t_0,\tau) \quad,\quad \mathcal{W}^s(\mathbf{x}_{0},t_0) = \textrm{argmin } M_p^{(f)}(\mathbf{x}_{0},t_0,\tau)
\label{min_LD_manifolds}
\end{equation}

\noindent where $\mathcal{W}^u$ and $\mathcal{W}^s$ are, respectively, the unstable and stable manifolds calculated at time $t_0$ and $\textrm{argmin}$ denotes the phase space coordinates $\mathbf{x}_0$ that minimize the function $M_p$. In addition, NHIMs at time $t_0$ can be calculated as the intersection of the stable and unstable manifolds:
\begin{equation}
\mathcal{N}(\mathbf{x}_{0},t_0) = \mathcal{W}^u(\mathbf{x}_{0},t_0) \cap \mathcal{W}^s(\mathbf{x}_{0},t_0) = \textrm{argmin } M_p(\mathbf{x}_{0},t_0,\tau)
\label{min_NHIM_LD}
\end{equation}

\noindent
It is important to point out here that the phase space location of the stable and unstable manifolds can be thus obtained in two ways. Firstly, we can extract them as ridges of the scalar function $|| \nabla M_p ||$ since manifolds are located at pòints where the function $M_p$ is non-differentiable. Once the manifolds are known, one can compute the NHIM at their intersection by means of a root search algorithm. The second method to recover the manifolds and their associated NHIM is by minimizing the function $M_p$ using a search optimization algorithm. This second procedure and some interesting variations are described in \cite{feldmaier2019}.

Notice that the LD definition given in Eq. (\ref{Mp_function}) implies that all initial conditions are integrated for the same time $\tau$. Recent studies have revealed, see e.g. \cite{junginger2017chemical,naik2019b,GG2019}, that computing fixed-time LDs, that is, integrating all initial conditions chosen on a phase space slice for the same integration time $\tau$, could give rise to issues related to the fact that some of the trajectories that escape the PES can go to infinity in finite time or at an increasing rate. The trajectories that show this behavior will give NaN values in the LD scalar field, hiding some regions of the phase space, and therefore obscuring the detection of invariant manifolds. In order to circumvent this problem we will apply in this work the approach that has been recently adopted in the literature \cite{junginger2017chemical,naik2019b,GG2019} known as variable integration time Lagrangian descriptors. In this methodology, LDs are calculated, at any initial condition, for the initial fixed integration time or until the trajectory of that initial condition leaves a certain phase space region $\mathcal{R}$ that we call the {\em interaction region}. Therefore, the total integration time in this strategy depends on the initial conditions themselves, that is $\tau(\mathbf{x}_0)$. In this variable-time formulation, the $p$-norm definition of LDs has the form:
\begin{equation}
M_p(\mathbf{x}_{0},t_0,\tau) = \int^{t_0 + \tau^{+}_{\mathbf{x}_0}}_{t_0 - \tau^{-}_{\mathbf{x}_0}} \sum_{i=1}^{n} |v_{i}(\mathbf{x}(t;\mathbf{x}_0),t)|^p \; dt \;,\quad p \in (0,1] \;.
\label{Mp_vt}
\end{equation}

\noindent
and, for a fixed integration time $\tau_0 > 0$, the total integration time is defined as:
\begin{equation}
\tau^{\pm}_{\mathbf{x}_{0}} = \min \left\lbrace \tau_0 \, , \, |t^{\pm}|_{\big| \mathbf{x}\left(t^{\pm}; \, \mathbf{x}_{0}\right) \notin \mathcal{R}} \right\rbrace \; ,
\end{equation}

\noindent
where $t^{+}$ and $t^{-}$ are the times for which the trajectory leaves the interaction region $\mathcal{R}$ in forward and backward time, respectively. For the analysis of the Caldera-type Hamiltonian in this work we have chosen:
\begin{equation}
\mathcal{R} = \left\lbrace \mathbf{x} = (x,y,p_x,p_y) \in \mathbb{R}^4 \; \big| \; |y| \leq 6 \right\rbrace \;.
\label{inter_reg}
\end{equation}

\noindent

To finish this section we will illustrate how variable integration time LDs can be used to detect the geometrical phase space structures, that is, the NHIMs and their stable and unstable invariant manifolds that characterize the dynamical matching phenomenon in the Caldera Hamiltonian system. In particular, we will focus on the extraction of the phase space structures for the dynamical system given in Eq. \eqref{eq3} using the model parameters described in Section \ref{sec.1}, and considering the unstretched ($\lambda = 1$) Caldera potential.
To compare the results obtained using LDs with those found in \cite{katsanikas2018} by means of other nonlinear dynamics techniques, we will analyze the phase space structures in the following Poincar\'e surfaces of section (SOSs):
\begin{eqnarray}
\mathcal{U}^{+}_{x,p_x} &=& \lbrace (x,y,p_x,p_y) \in \mathbb{R}^4 \;|\; y = 1.88409 \; ,\; p_y > 0 \;,\; E = 29 \rbrace \\
\mathcal{V}^{+}_{x,p_x} &=& \lbrace (x,y,p_x,p_y) \in \mathbb{R}^4 \;|\; y = 0 \; ,\; p_y > 0 \;,\; E = 30 \rbrace
\label{psos_defs}
\end{eqnarray}

\begin{figure}[htbp]
	\centering
	A)\includegraphics[scale=0.26]{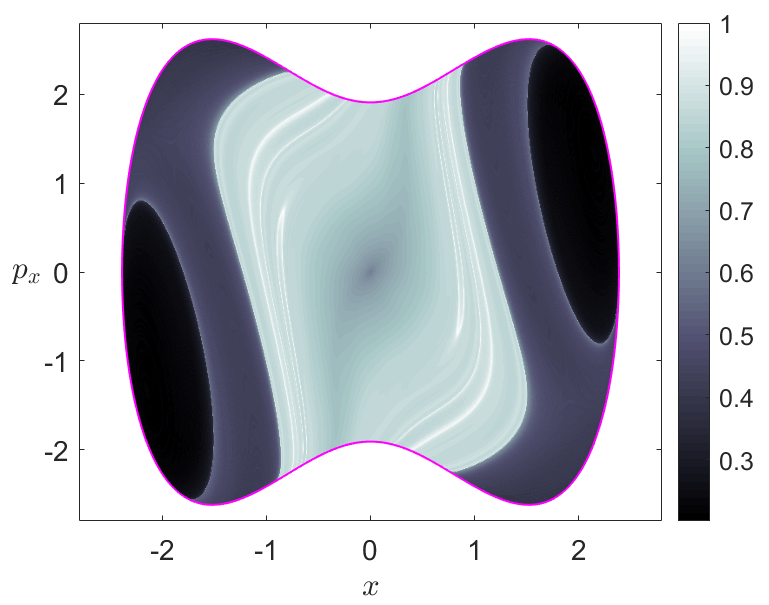}
	B)\includegraphics[scale=0.26]{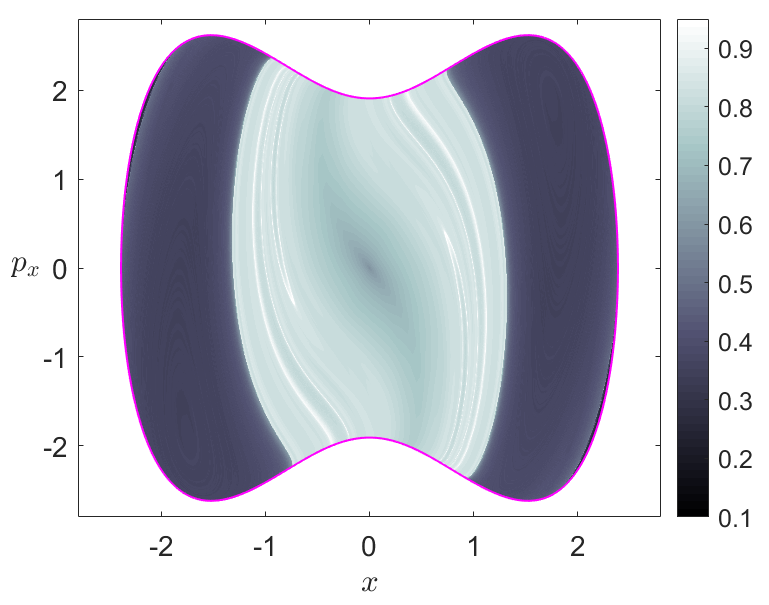}
	C)\includegraphics[scale=0.26]{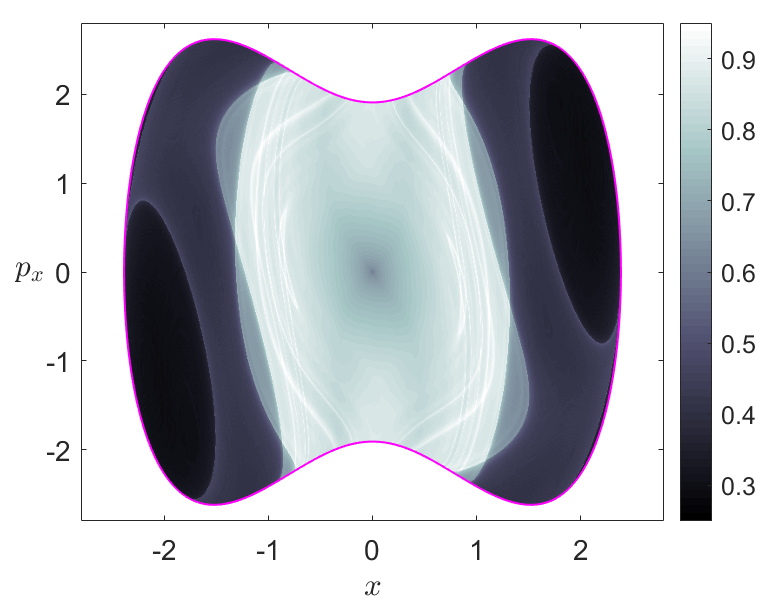}
	\caption{Computation of variable-time LDs in the Poincar\'e SOS $\mathcal{U}^{+}_{x,p_x}$ using $\tau = 4$ and $p = 1/2$. A) Forward integration LDs; B) Backward integration LDs; C) The sum of forward and backward LDs. The energy boundary is depicted in magenta.}
	\label{fig:LD_tau4}
\end{figure}

\begin{figure}[htbp]
	\centering
	A)\includegraphics[scale=0.38]{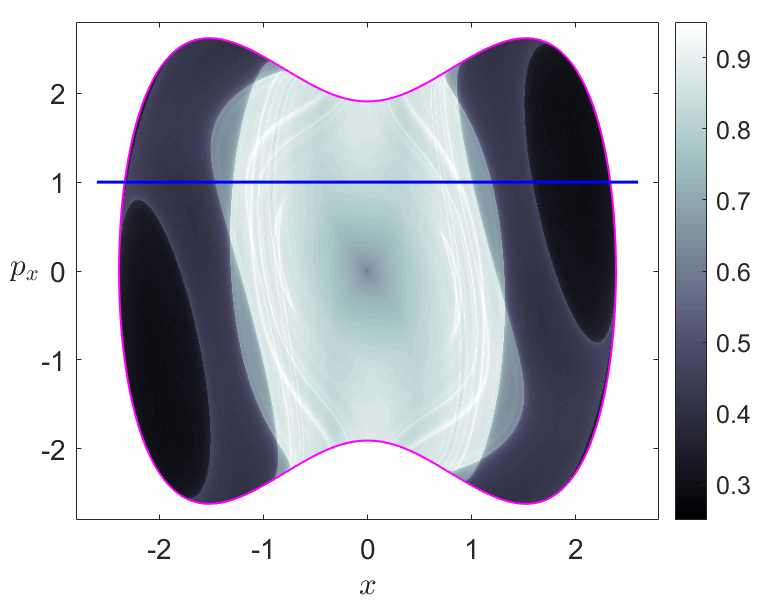}
	B)\includegraphics[scale=0.38]{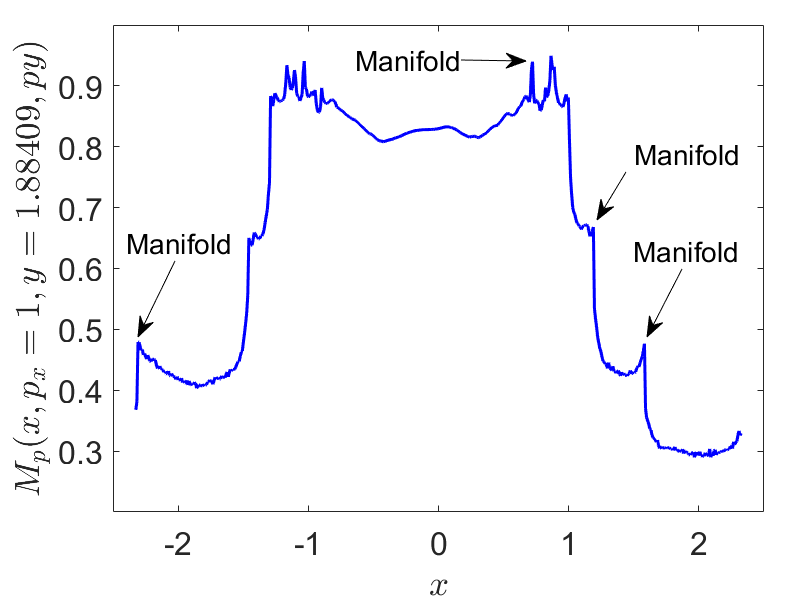}
	\caption{Detection of stable and unstable manifolds at phase space points where the LD scalar function is non-differentiable. A) Variable-time LDs calculated on the Poincar\'e SOS $\mathcal{U}^{+}_{x,p_x}$ using $\tau = 4$ and $p = 1/2$; B) Value of LDs along the line $p_x = 1$.}
	\label{fig:LD_tau4_maniDetect}
\end{figure}

\begin{figure}[htbp]
	\centering
	A)\includegraphics[scale=0.4]{LD_lambda_1_y_188409_E_29_tau_4.png}
	B)\includegraphics[scale=0.4]{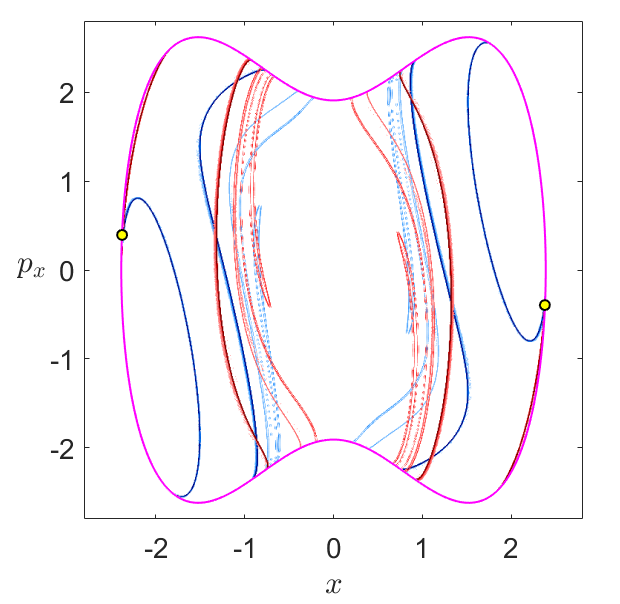} 
	C)\includegraphics[scale=0.4]{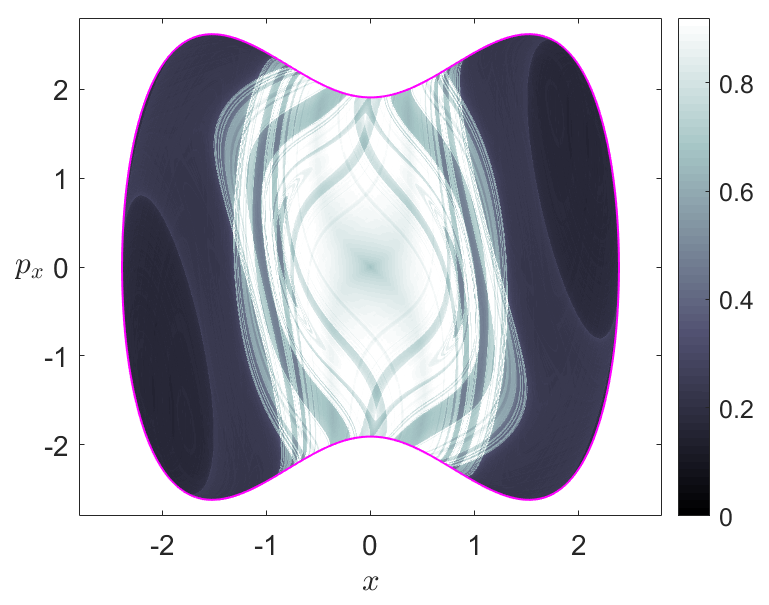}
	D)\includegraphics[scale=0.4]{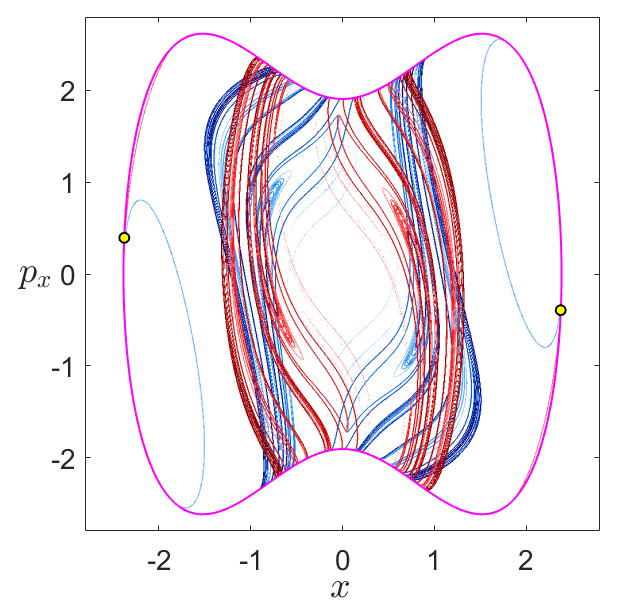}
	\caption{On the left column, LDs calculated on the SOS $\mathcal{U}^{+}_{x,p_x}$ using: A) $\tau = 4$; C) $\tau = 8$. On the right column, the invariant stable (blue) and unstable (red) manifolds extracted from the gradient of the scalar function $M_p$. We have also marked with yellow dots the location of the unstable periodic orbits of the upper index-1 saddles and the magenta curve represents the energy boundary.}
	\label{fig:LD_mani_extract}
\end{figure}

\noindent
We begin our analysis with the SOS $\mathcal{U}^{+}_{x,p_x}$, and we choose a small integration time $\tau	 = 4$. Once we have fixed the phase space slice where we want to compute LDs, we select a grid of initial conditions and, after discarding those that are energetically unfeasible, we integrate the remaining conditions both forward and backward in time, and compute LDs using the definition in Eq. \eqref{Mp_vt} with $p = 1/2$  for the whole fixed integration time or until the trajectory leaves the interaction region $\mathcal{R}$ in Eq. \eqref{inter_reg}, whichever happens first. The result is that if we plot the LDs values obtained from the forward/backward integration, the scalar field will reveal the stable/unstable manifolds in the SOS under consideration. Moreover, if we plot the combined sum of forward and backward integration, the method highlights both stable and unstable manifolds simultaneously. This is shown in Fig. \ref{fig:LD_tau4}, where the values of LDs for forward/backward integration is displayed in panel A)/B) and the combination of both is depicted in C). We can clearly see that the manifolds are detected at points where the LD scalar function is non-differentiable. To demonstrate this mathematical property, we represent in Fig. \ref{fig:LD_tau4_maniDetect} the values taken by the LD function calculated on $\mathcal{U}^{+}_{x,p_x}$ along the line $p_x = 1$. Notice the jumps in the values of the function, which indicate non-differentiability by means of  large gradient values. Therefore, we can directly extract the invariant stable and unstable manifolds in the SOS from the gradient, that is, using $||\nabla M_p||$. This is illustrated in Fig. \ref{fig:LD_mani_extract} for the SOS $\mathcal{U}^{+}_{x,p_x}$ where two different values for the integration time have been used to compute LDs, in particular $\tau = 4$ and $\tau = 8$. It is important to note here the crucial role that the integration time $\tau$ plays when it comes to revealing the invariant manifolds in phase space. As shown in Fig. \ref{fig:LD_mani_extract}, when we increase the value for the integration time, richer and more complex details of the underlying geometrical template of phase space structures are unveiled. This behavior is expected, since an increase of the integration time would imply incorporating more information about the past and future dynamical history of particle trajectories in the computation of LDs. This means that $\tau$ is intimately related to the time scales of the dynamical phenomena that take place in the model under consideration and thus, it is a parameter that is problem-dependent. Consequently, there is no general ``golden'' rule for selecting its value for exploring phase space, and thus it is usually selected from the information obtained by performing several numerical experiments. One needs to always bare in mind that there is a compromise between the complexity of the structures that one would like to reveal to explain a certain dynamical mechanism, and the interpretation of the intricate manifolds displayed in the LD scalar output. 

\begin{figure}[htbp]
	\centering
	A)\includegraphics[scale=0.39]{LD_lambda_1_y_188409_E_29_tau_8.png} 
	B)\includegraphics[scale=0.4]{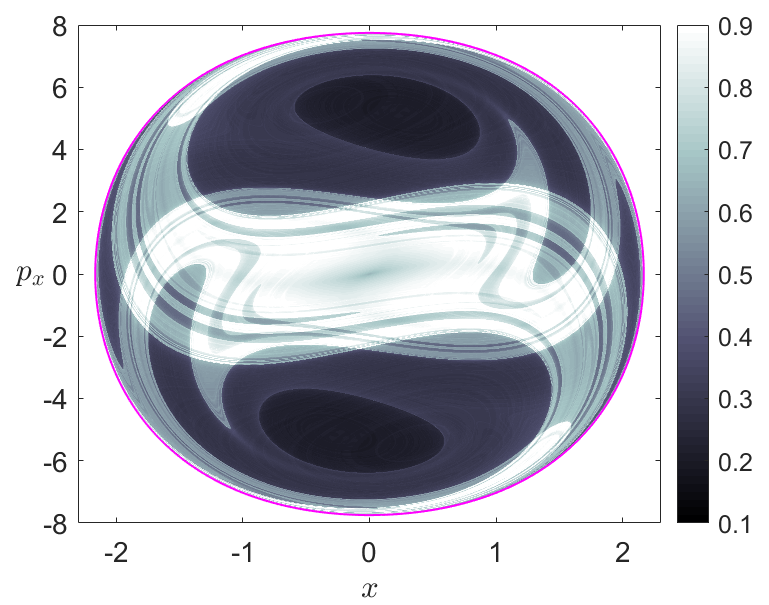}
	C)\includegraphics[scale=0.4]{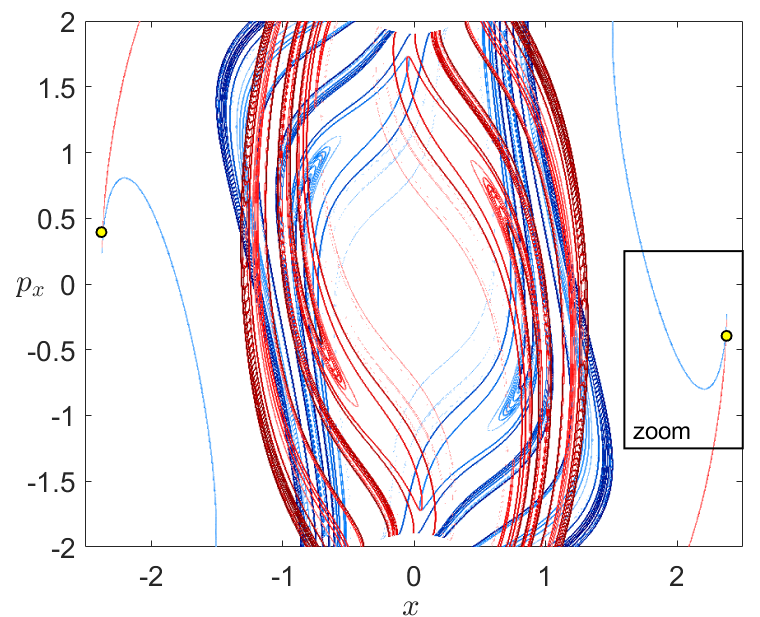}\hspace{.3cm}
	D)\includegraphics[scale=0.4]{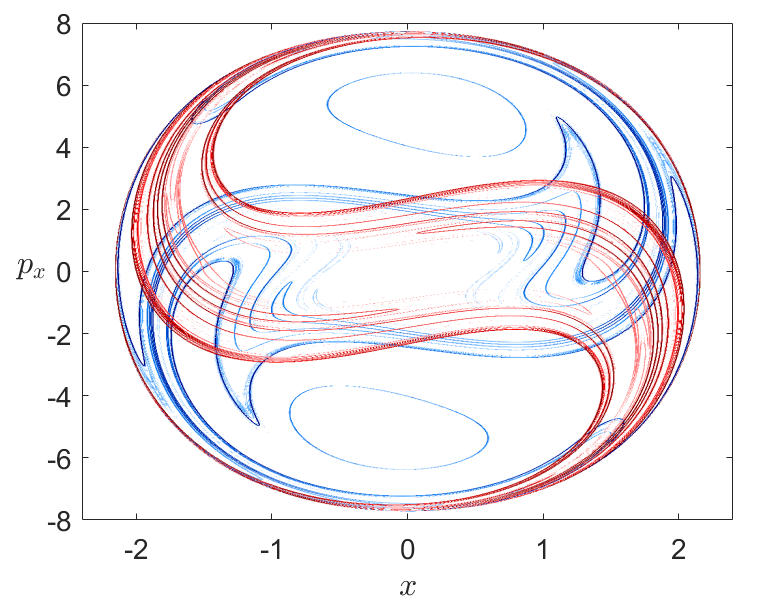}
	E)\includegraphics[scale=0.57]{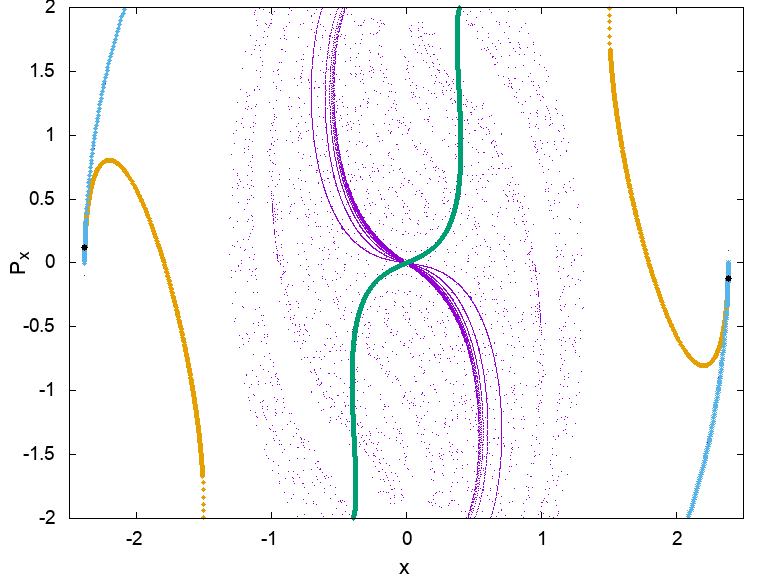}\hspace{.3cm}
	F)\includegraphics[scale=0.57]{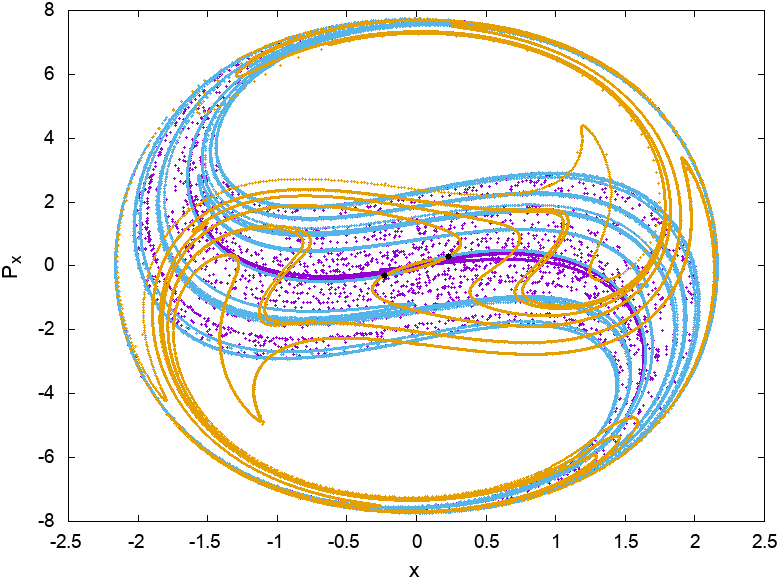}
	\caption{On the left column: A) LDs calculated on the SOS $\mathcal{U}^{+}_{x,p_x}$ using $\tau = 4$; C) invariant stable (blue) and unstable (red) manifolds extracted from the gradient of the scalar function $M_p$; E) and F) Unstable (cyan) and stable
		(orange) invariant manifolds of the periodic orbits of the
		two upper saddles, that are also represented by two black points. We also depict the invariant unstable (violet) and stable (green) manifolds of the family of periodic orbits of the central minimum. On the right column we perform the same analysis but for the Poincar\'e SOS $\mathcal{V}^{+}_{x,p_x}$, where LDs have been calculated using an integration time $\tau = 6$. It is important to remark that the invariant manifolds shown in E) and F) have been computed by means of classical nonlinear techniques, see \cite{katsanikas2018}.}
	\label{fig:LD_PSec_comp}
\end{figure}

As a final remark to complete the analysis of this example on how the method of Lagrangian descriptors is applied to extract the geometrical template of invariant manifolds in a high-dimensional phase space by means of looking at low-dimensional slices, there is a key point that needs to be highlighted and that demonstrates the real potential of LDs with respect to other classical techniques from nonlinear dynamics. In Figs. \ref{fig:LD_mani_extract} and \ref{fig:LD_PSec_comp} we have extracted from the gradient of the $M_p$ function the stable and unstable manifolds on the Poincar\'e sections $\mathcal{U}^{+}_{x,p_x}$ and $\mathcal{V}^{+}_{x,p_x}$ respectively. Using LDs we can obtain \textit{all} the manifolds coming from \textit{any} NHIM in phase space \textit{simultaneously}. This is of course a tremendous advantage in comparison to the classical approach used to compute stable and unstable manifolds, which relies on the individual location of the NHIMs in phase space and, for every NHIM, globalize the manifolds separately taking into account the crucial information provided by the eigendirections. Consequently, the application of LDs offers the capability of recovering \textit{all} the relevant phase space structures in one \textit{shot} without having to study the local dynamics about equilibrium points of the dynamical system. To validate that the structures extracted from the gradient of LDs correspond to the stable and unstable manifolds present in the phase space of the Caldera Hamiltonian, we have compared them in Fig. \ref{fig:LD_PSec_comp} with the invariant manifolds obtained by means of classical nonlinear dynamics techniques to calculate periodic orbits, see \cite{katsanikas2018} for more details.

\section{Numerical Results}
\label{sec.2}

In this section we compute Lagrangian descriptors with $\tau = 4$ in order to study the phase space structures close to the UPOs associated with the upper index-1 saddles. Our goal is to reveal the phase space mechanism that determines the existence or nonexistence of dynamical matching in the Caldera PES. For this purpuse we use the Poincar\'e surfaces of section defined in Eq. \eqref{psos_defs}, which were  also used in \cite{katsanikas2018}. This analysis is carried out for different values of $\lambda$. Our goal is to understand how LDs are capable of detecting the dynamical matching mechanism. This section is divided into two subsections. In the first part we describe how the method of LDs succeeds in the detection of dynamical matching, and the second subsection presents the properties and advantages of this methodology.

\subsection{The detection of Dynamical Matching}

The phenomenon of dynamical matching refers to the lack of a mechanism that would enable transport of trajectories from the region of the upper saddles to the central area of the Caldera. As we know, trajectories  with initial conditions on the invariant manifolds of unstable periodic orbits move  away from the periodic orbit (unstable manifold)  or approach the periodic orbit (stable manifold). A mechanism  that  could be responsible for the transport of  trajectories from the region of the upper saddles  to the central area of the Caldera, would be heteroclinic intersections  of the unstable invariant manifolds of the unstable periodic orbits of the upper saddles  with the stable manifolds of the unstable periodic orbits that exist in the central area. We will show that the non-existence or existence of  this mechanism determines if we  have dynamical matching or not. For this reason, we compute the invariant manifolds for different values of $\lambda$ starting from $\lambda=1$ to zero in order to find the values of $\lambda$ that correspond to  dynamical matching  and trapping.

\begin{enumerate}

\item \underline{\textbf{Dynamical matching:}} The  gap  in  Fig.\ref{fig1}  (for $\lambda=0.8$)  indicates  that we have no interaction (heteroclinic intersections) of the  unstable invariant manifold of the  periodic orbits associated with the upper saddle with the central area  and this means that we have no mechanism of transport of trajectories from one region to the other. Consequently,  we have in this case the phenomenon of dynamical matching, the trajectories that have initial conditions on the dividing surfaces of the periodic orbits of upper saddles go straight across the Caldera and they exit via the lower opposite saddle as we know  from  previous papers (\cite{katsanikas2018}, \cite{katsanikas2019}). An example of this is given in Fig. \ref{fig1} for $\lambda=0.8$. As we can see in this figure  we choose an initial conditions  (circle) inside the region of the unstable invariant manifold of the unstable periodic orbits of upper saddles. If we integrate backward the initial condition that corresponds to the circle the resulting trajectory exits via the region of the upper saddle. If we integrate it forward the resulting trajectory goes straight across the caldera and exits via the lower opposite  saddle. This means that the trajectory comes from the region of the upper saddle and it exhibits the phenomenon of dynamical matching. This gap  decreases in size as we decrease the stretching parameter $\lambda$ until we reach a  critical value of $\lambda$. \\

\item \underline{\textbf{The critical value}:} In Fig. \ref{fig1} we observe for  $\lambda=0.778$ (middle row of figures)  the unstable manifolds of the periodic orbits of upper saddles start to interact with the stable manifolds of the unstable periodic orbits of the central area, resulting in  heteroclinic connections and  forming lobes between them. These lobes are very narrow and cannot be distinguished initially  as we can see in  Fig.\ref{fig1}. In order to observe  these lobes  we magnify the region of the  upper saddles, for example the region of the upper right saddle in Fig.\ref{fig1}. When we magnify these regions, we see  the  heteroclinic connections and the lobes between  the unstable invariant manifolds of the unstable periodic orbits of upper saddles and the stable manifolds of the unstable periodic orbits that exist in the central area. These lobes are responsible for the trapping of the trajectories that come from the region of the upper saddles to the central area. This can be checked very easily. We depict two initial conditions in  Fig.\ref{fig1} for  $\lambda=0.778$, one inside the lobe (the diamond) and other one outside the lobe (the circle) but inside the region of the unstable manifold of the unstable periodic orbit of upper saddle. If we integrate backward the two initial conditions, we see that the corresponding trajectories come from the region of the right upper saddle because  they exit via the region of the right upper saddle. But if we integrate forward the initial condition, that is inside the lobe, the corresponding trajectory is trapped and after a long time exits through the region of the  opposite lower saddle. On the contrary, the trajectory that corresponds to the  other  initial condition is not trapped and go straight across to the exit from the caldera. This means that the initial conditions in the lobes between the unstable invariant manifolds of the unstable periodic orbits associated with the upper saddles and the stable invariant manifolds of the unstable periodic orbits of the central area are responsible for the trapping of the trajectories that come from the region of the upper saddles. This is the first value of $\lambda$ for which we find interaction between the unstable invariant manifolds of unstable periodic orbits, associated with the upper saddles, with the central area. This means that this  is a critical value of the  stretching parameter for the non-existence of  dynamical matching, as we have observed in a previous paper \cite{katsanikas2019}). \\

\item \underline{\textbf{Trapping}:} Now if we decrease the value of $\lambda$, starting from the critical value, we have again interaction of the unstable invariant manifolds of unstable periodic orbits of upper saddles  with the central area. We have again lobes between the unstable invariant manifolds of the unstable periodic orbits with the stable invariant manifolds of the unstable periodic orbits that exist in the central region as we can see for example for $\lambda=0.7$ in Fig.\ref{fig1}. This means that we have again trapping for values of $\lambda$ lower than the critical value.

\end{enumerate}

\subsection{Properties and advantages of the method of Lagrangian Descriptors.}

\begin{figure}[htbp]
	\centering
	A)\includegraphics[scale=0.25]{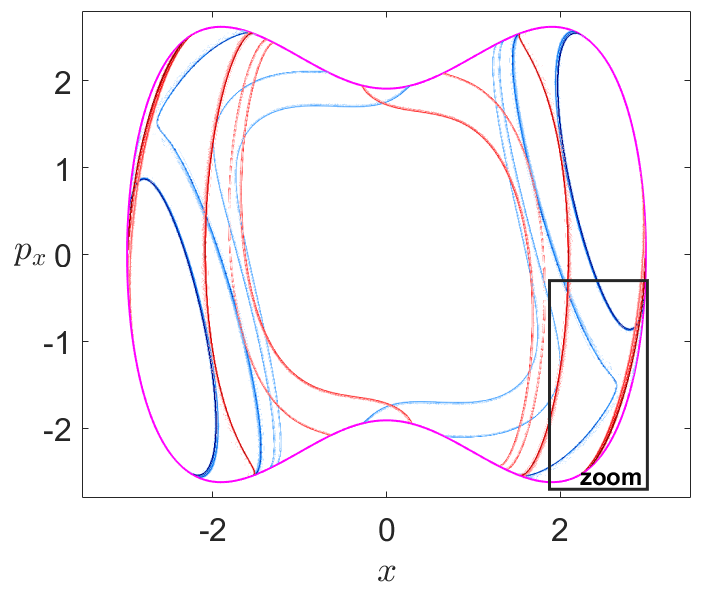}
	B)\includegraphics[scale=0.25]{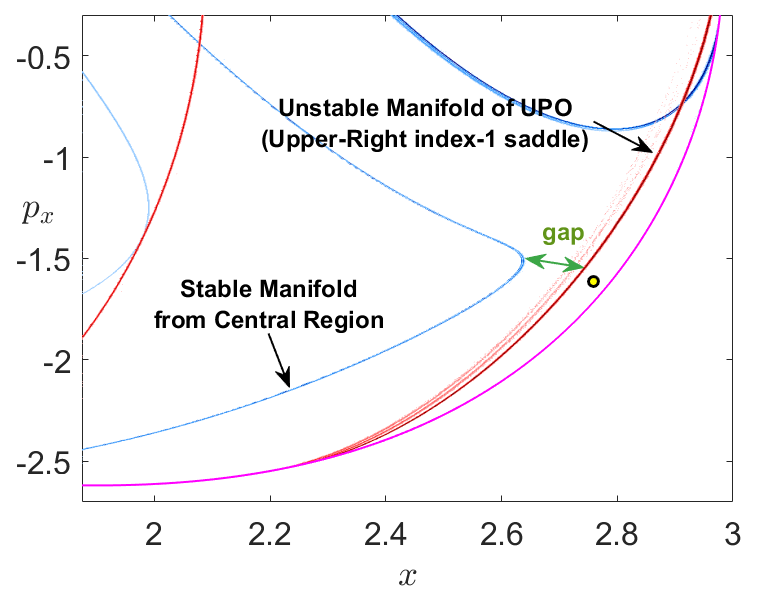}
	C)\includegraphics[scale=0.25]{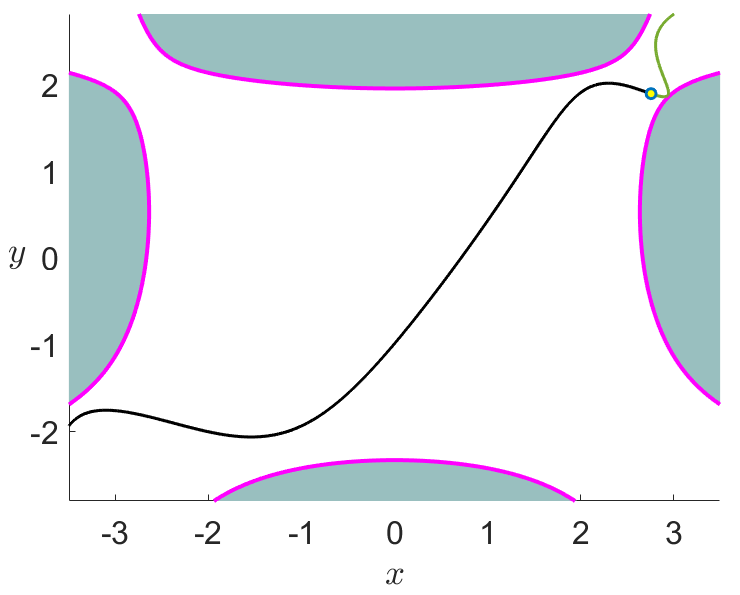}\\
	D)\includegraphics[scale=0.25]{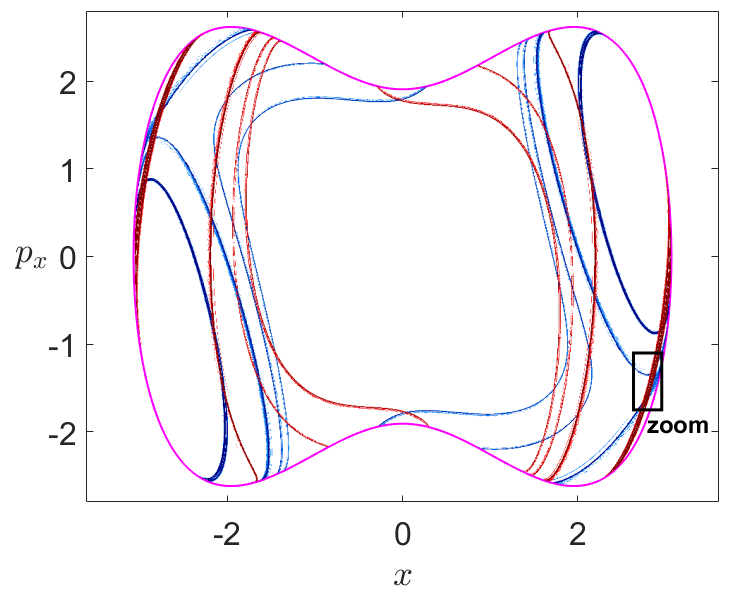}
	E)\includegraphics[scale=0.25]{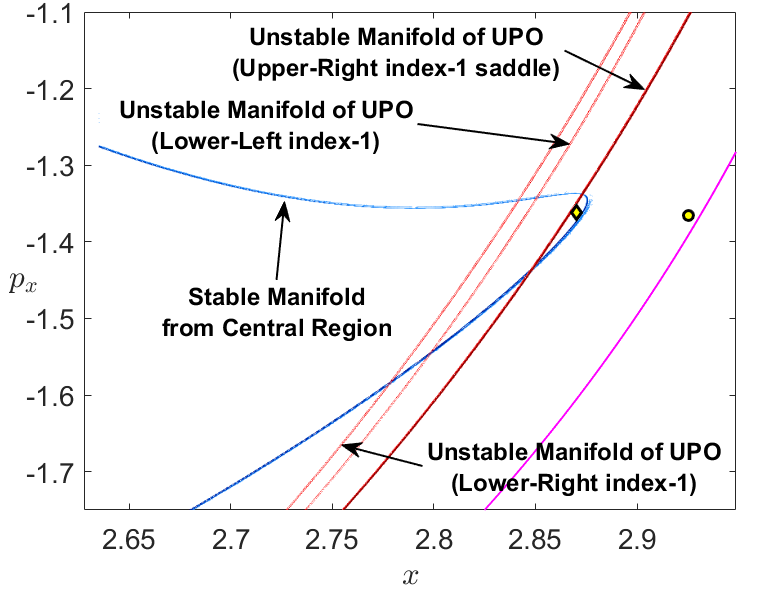}
	F)\includegraphics[scale=0.25]{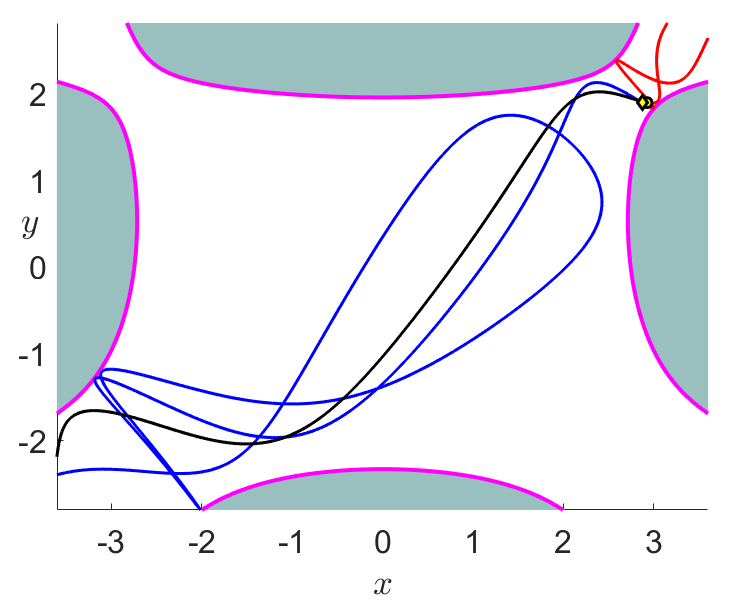}\\
	G)\includegraphics[scale=0.25]{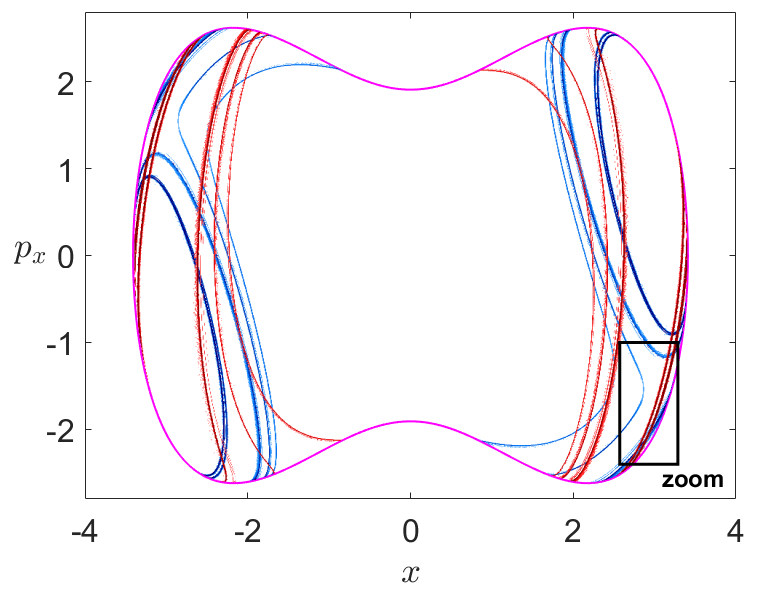}
	H)\includegraphics[scale=0.25]{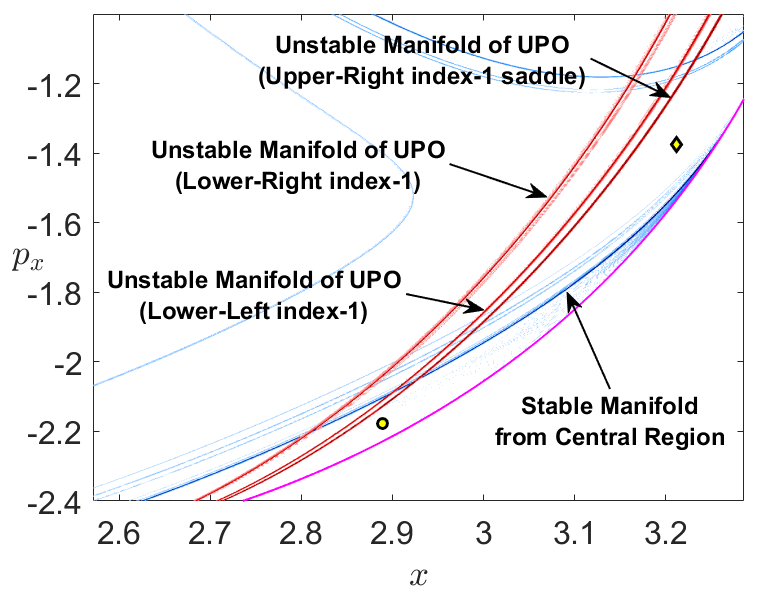}
	I)\includegraphics[scale=0.25]{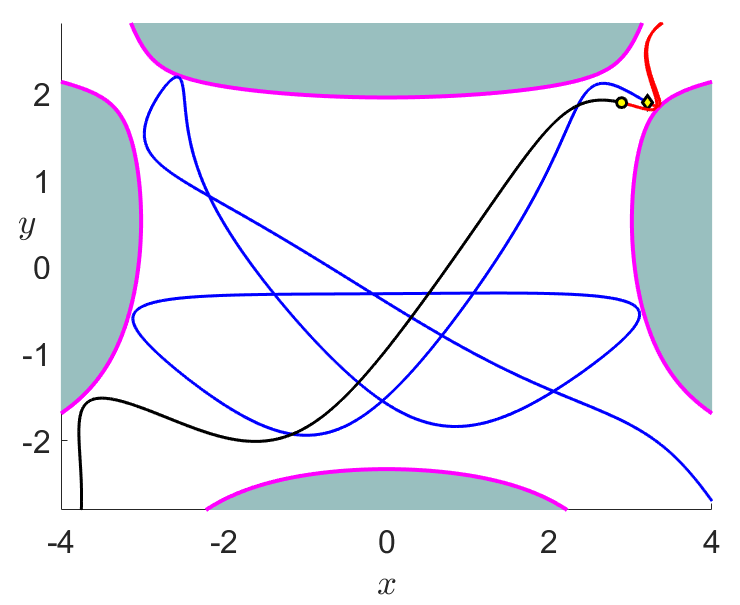}\\
	\caption{(First column) Phase space structures close to the UPOs associated with the upper saddles; (Second column) Zoom of the phase space region indicated in the figures of the first column. The stable (blue) and unstable (red) manifolds have been revealed by applying LDs with $\tau = 4$; (Third column) projection onto configuration space of the trajectories that start from the initial conditions marked in the second column as a circle and a diamond. Black and blue curves correspond to forward time integration, while red and green are for backward integration. Panels (a)-(c) are for the stretching parameter $\lambda = 0.8$, (d)-(f) use $\lambda = 0.778$, and (g)-(i) correspond to $\lambda = 0.7$.}
	\label{fig1}
\end{figure}

\begin{figure}[htbp]
\centering
A)\includegraphics[angle=0,width=8cm]{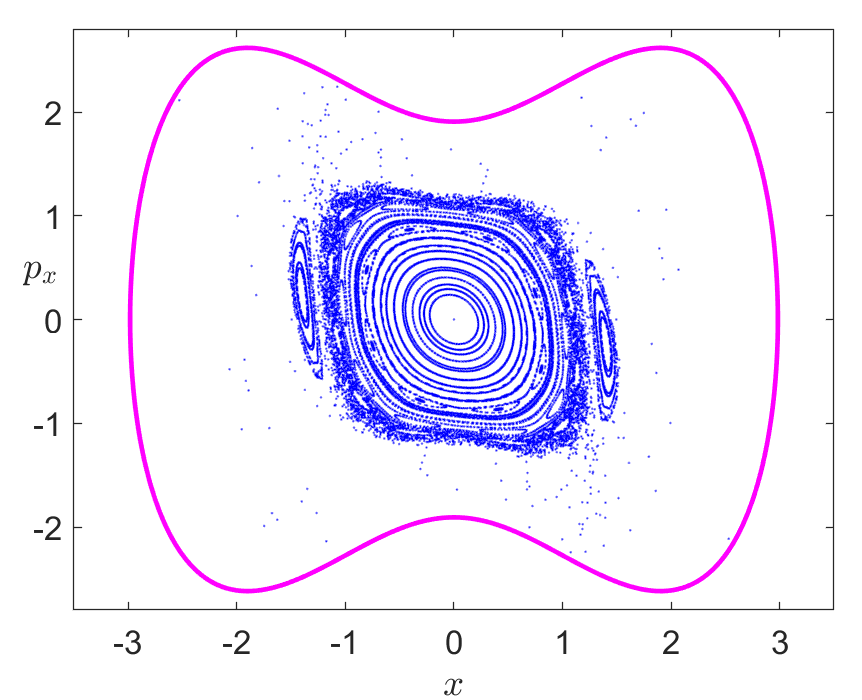}
B)\includegraphics[angle=0,width=8cm]{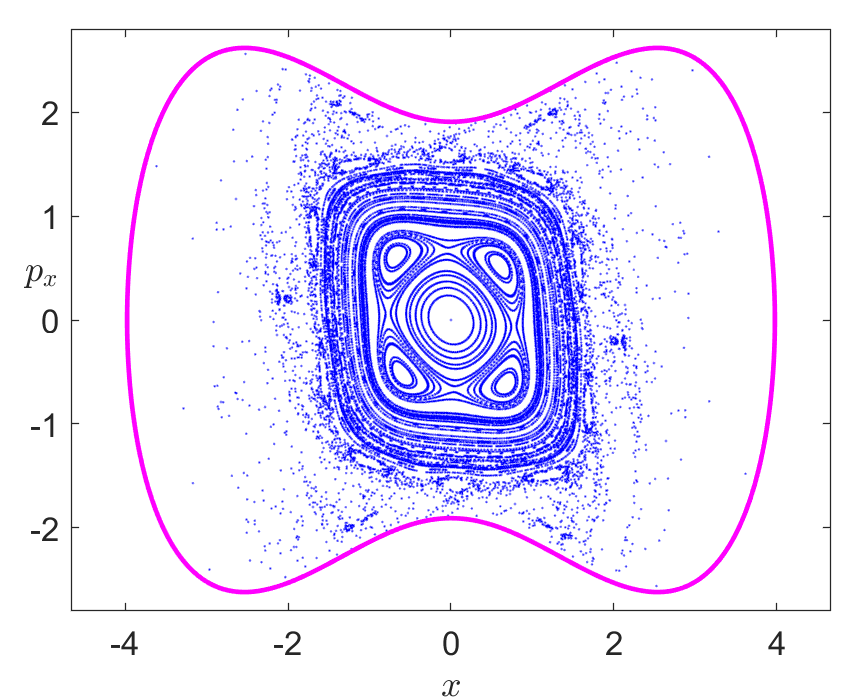}
C)\includegraphics[angle=0,width=8cm]{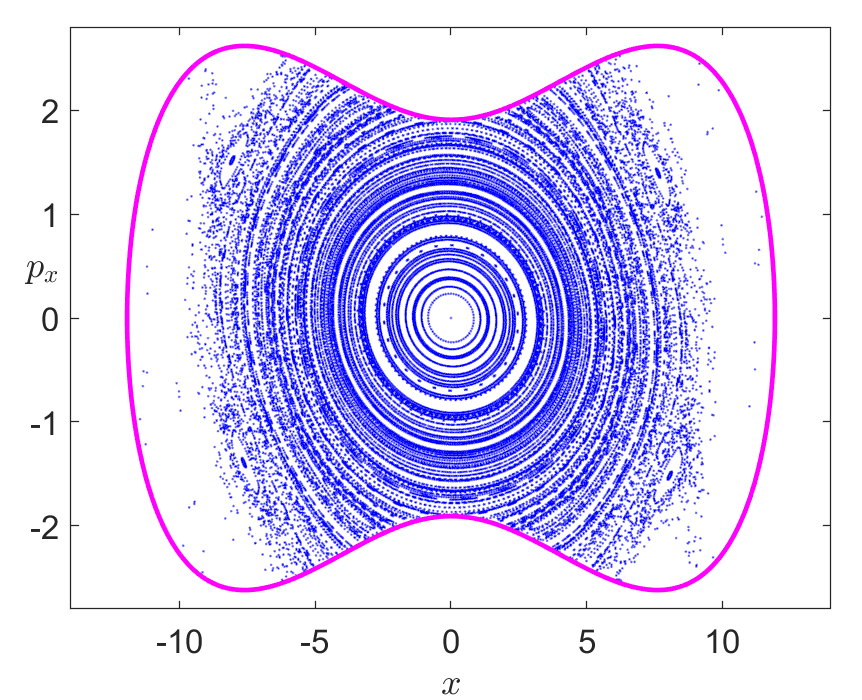}
\caption{Phase space close to the unstable periodic orbits associated with the upper saddles using the Poincar{\'e} surface of section $y=1.884090$ with $p_y>0$ at energy $E=29$ for the stretching parameter: A) $\lambda=0.8$; B) $\lambda=0.6$; C) $\lambda=0.2$.}
\label{fig2}
\end{figure}

\begin{figure}[htbp]
\centering
\includegraphics[scale=0.45]{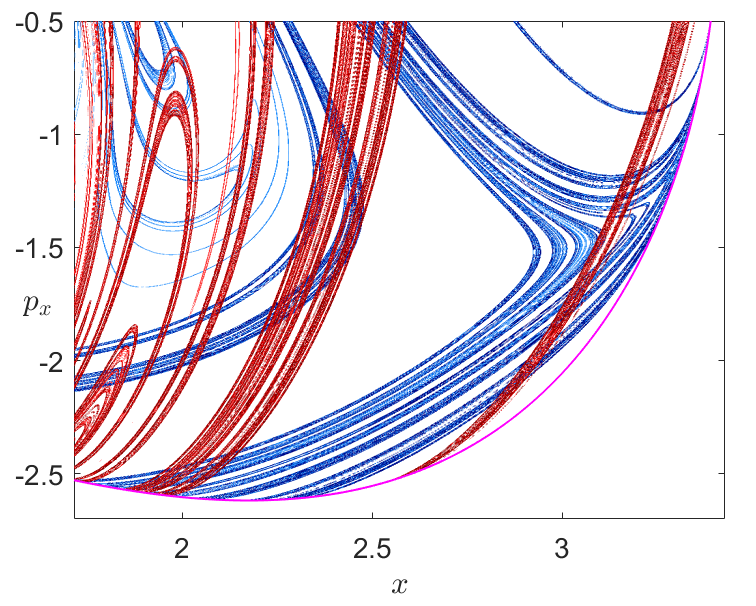}
\caption{Phase space stable (blue) and unstable (red) manifolds extracted from Lagrangian descriptors close to the unstable periodic orbits associated with the upper index-1 saddles. The computation has been carried out using $\tau = 15$ for the Hamiltonian system with energy $E = 29$ and stretching parameter $\lambda = 0.7$ in the Poincar{\'e} section $y = 1.884090$ with $p_y > 0$.}
\label{fig3b}
\end{figure}

In this subsection we describe three different properties and advantages of the method of Lagrangian descriptors for the detection of dynamical matching:

\begin{enumerate}
    \item \underline{\textbf{Accuracy:}} An important advantage of  Lagrangian descriptors is that they provide a more accurate approximation of the critical value of the stretching parameter for the transition from the case of the dynamical matching to the case of the non-existence  of the dynamical matching, than the approximations that are obtained  from other methods such as  dividing surfaces. For example in this paper,  the critical value $\lambda=0.778$, that we computed using Lagrangian descriptors,  is a little larger than the critical value $\lambda=0.72$, that is computed using dividing surfaces (see \cite{katsanikas2019}). The trapping of the trajectories is obtained for values of the stretching parameter below the critical value, depend on the formation of a narrow lobe, see Fig. 7, between the unstable invariant manifolds of the unstable periodic orbits of the upper saddles and the stable manifolds of the unstable periodic orbits that exist in the central area, as we explained  earlier. This narrow lobe can be very easily  identified using Lagrangian descriptors because we can see directly which part of the phase space can be responsible for the trapping and transport of the trajectories   from the region of the upper saddles to the central area of the Caldera. But if we use the dividing surfaces we are constrained to identify the  phenomenon of  dynamical matching in the configuration space without knowing the structure of the phase space and if there is a region of the phase space that is responsible for the trapping of the trajectories in the central area of the Caldera. This means that it depends  on the  sampling of the dividing surface whether or not we will detect the   phenomenon of dynamical matching. For the case of the critical value we have only very few trajectories that are trapped inside a narrow lobe and this  makes  it very difficult for these trajectories to be included in the sampling of the dividing surfaces. For this reason, we can identify the critical value with more accuracy  using Lagrangian descriptors. \\[.1cm]
    
\item \underline{\textbf{The integration time $\tau$}:} A crucial quantity for the detection of   dynamical matching is the time $\tau$ of the computation of Lagrangian descriptors. In all cases we used  $\tau=4$ because we could see all the appropriate geometrical structures and specifically  the invariant manifolds of the unstable periodic orbits of the upper saddles and  the invariant manifolds of the unstable periodic orbits of the central area. This could allow us to see directly if we have a gap or lobe (dynamical matching or trapping)  between the unstable invariant manifolds of the unstable periodic orbits associated with the  upper saddles and the stable manifolds of the unstable periodic orbits that exist in the central area. For values of $\tau$ less than $4$ we could not see, in many cases,  the invariant manifolds  from the central area of the Caldera.  On the contrary, for larger values of $\tau$ we could see more structures but it was very difficult to detect the appropriate lobes that were responsible for the non-existence of the dynamical matching. For example we identify for $\lambda=0.7$  and   $\tau=4$ (see Fig. \ref{fig1}) the non-existence of the dynamical matching because of the lobe  between the unstable invariant manifolds of the unstable periodic orbits associated with the upper saddles and the stable manifolds of the unstable periodic orbits that exist in the central area. But, if we use  large values for $\tau$, as  for example $\tau=15$ (Fig.\ref{fig3b}), we have many returns of the invariant manifolds and it is not obvious which lobe is responsible for the trapping of the trajectories that come from the region of the upper saddles. This means that increasing the time $\tau$,  we increase the complexity of the  figures and it is very difficult to detect the non-existence of the dynamical matching. If we decrease  the time $\tau$ less than $4$  we cannot also identify the existence or not of the dynamical matching  because some of the geometrical structures from the central area   do not exist in the figures. There is a critical value for $\tau$ that is sufficient to see the appropriate geometrical structures (invariant manifolds from the region of the upper saddles and central area) and to detect lobes and gaps between them  but also it is not so large as to increase  the  complexity of the figures. In our paper this value is $\tau=4$. \\[.1cm]

\item \underline{\textbf{The increase of Trapping:}} Using the method of Lagrangian descriptors we can predict the increase of trapping as we decrease the stretching parameter. As we decrease the $\lambda$ parameter we approach the integrable case of our system. The integrable case of our system corresponds to $\lambda=0$. In this case there is no x coordinate in the expression for the caldera PES  and our system has only one degree of freedom, and  it is therefore integrable. This is the reason as we can see in  Fig.\ref{fig2} the ordered region around the central stable periodic orbit increases, as we decrease the $\lambda$ parameter, decreasing the ratio of the free space for  the invariant manifolds of the unstable periodic orbits  to the permitted energy region (that is indicated by magenta color in  Fig. \ref{fig2}). Consequently, the stable  invariant manifolds of the unstable periodic orbits, that exist in the central area,  open more and more to the edge of the permitted space forming larger lobes with the unstable invariant manifolds of the unstable periodic orbits associated with the upper saddles. We can see this for example if we compare the lobes between the case for $\lambda=0.778$ and $\lambda=0.7$ (in Fig. \ref{fig1}). The increasing size of lobes can explain the increase of the trapping of trajectories in the central area, as we decrease the $\lambda$ parameter,  which was also   observed in  a  previous paper \cite{katsanikas2019}.  
\end{enumerate}

\section{Conclusions}

In this work we have used the method of Lagrangian descriptors to detect the dynamical matching mechanism in a Caldera-type Hamiltonian system stretched in the $x$-direction, and our analysis has helped us to develop a deeper understanding of the dynamical origin of this phenomenon in phase space. The results we have found in this study are:

\begin{enumerate}
\item  Lagrangian descriptors can easily detect the gap between the unstable invariant manifolds of the upper index-1 saddles and the stable manifolds of the unstable periodic orbits that exist in the central area. This gap corresponds to  dynamical matching and is a consequence of the non-existence of a heteroclinic connection in phase space. \\

\item The detection of dynamical matching can be carried out only by means of the computation of LDs, allowing us to avoid the use of dividing surfaces, classical methods for finding periodic orbits, the use of Poincar{\'e} sections and the separate  computation of the invariant manifolds on Poincar{\'e} sections. This means that this method is faster and can be implemented in all cases even in systems with many escapes in which the computation of periodic orbits using classical methods and the use of dividing surfaces is difficult. \\

\item Lagrangian descriptors can detect not only the non-existence of dynamical matching but also the specific regions of the phase space that are responsible for this type of behavior. We can easily see using Lagrangian descriptors the interaction of the unstable manifolds of the unstable periodic orbits of the upper saddles with the stable manifolds of the unstable periodic orbits of the central area. Then we can identify which lobes between the unstable manifold of the unstable periodic orbits of upper saddles and  the stable manifolds of the unstable periodic orbits of the central area are responsible for the trapping of the trajectories. We can also predict if the intensity of the phenomenon of trapping in the central area of the Caldera  will be small or large from the size of the lobes. This gives us a deeper understanding of the origin of this phenomenon. \\

\item For the detection of dynamical matching the method of Lagrangian descriptors is more accurate than the sampling of dividing surfaces. This is because this mechanism may involve only a few special trajectories that could easily be missed in a sampling procedure. In particular, these trajectories come from the region of the upper saddles and are trapped in the central area of the Caldera. Narrow lobes between  the unstable manifolds of the unstable periodic orbits of the upper saddles with the stable manifolds of the unstable periodic orbits of the central area are responsible for this trajectory behaviour. \\[.1cm]

\item The detection of dynamical matching by means of Lagrangian descriptors is very sensitive to the value chosen for the integration time $\tau$ to compute LDs. By numerical experiments and inspection one can easily find a suitable value so that the method clearly reveals the relevant invariant manifolds in the region of the upper index-1 saddles and the central area of the Caldera, allowing for the detection of lobes and gaps between manifolds. As we have pointed out, the selection of $\tau$ is a relevant step in the process, since for large integration time values, the complexity of the phase space structures recovered by this technique would make the interpretation of figures a difficult task. This phenomenon is illustrated in Fig. \ref{fig3b}. 

\end{enumerate}

\nonumsection{Acknowledgments:} 

The authors acknowledge the support of EPSRC Grant no. EP/P021123/1 and Office of Naval Research Grant No. N00014-01-1-0769.

\bibliographystyle{ws-ijbc}
\bibliography{caldera2c}

\end{document}